\documentclass[10pt,
nofootinbib,reprint,aps,prd]{revtex4-1}
\usepackage{amsfonts,amsmath,amssymb,mathrsfs}
\usepackage{mathtools,mathptmx,array}
\usepackage{graphics,graphicx}
\usepackage{subfigure,color}
\usepackage[colorlinks=true,linkcolor=blue,citecolor=blue,urlcolor=blue]{hyperref}

\def\be{\begin{equation}}
\def\ee{\end{equation}}
\def\bea{\begin{eqnarray}}
\def\eea{\end{eqnarray}}
\def\nn{\nonumber}
\def\p{\partial}
\def\cA{\mathcal{A}}

\begin{document}


\title{Aspects of the dyonic Kerr-Sen-AdS$_4$ black hole and its ultraspinning version}

\author{Di Wu$^{1,2}$}
\email{wdcwnu@163.com}

\author{Shuang-Qing Wu$^{2}$}
\email{Corresponding author. sqwu@cwnu.edu.cn}

\author{Puxun Wu$^{1}$}
\email{Corresponding author. pxwu@hunnu.edu.cn}

\author{Hongwei Yu$^{1}$}
\email{hwyu@hunnu.edu.cn}

\affiliation{$^{1}$Department of Physics and Synergetic Innovation Center for
Quantum Effect and Applications, Hunan Normal University, Changsha, Hunan 410081,
People's Republic of China \\
$^{2}$College of Physics and Space Science, China West Normal University, Nanchong,
Sichuan 637002, People's Republic of China}

\date{\today}

\begin{abstract}
We explore some (especially, thermodynamical) properties of the dyonic Kerr-Sen-AdS$_4$ black
hole and its ultraspinning counterpart, and check whether or not both black holes satisfy the
first law and Bekenstein-Smarr mass formulas. To this end, new Christodoulou-Ruffini-like
squared-mass formulae for the usual dyonic Kerr-Sen-AdS$_4$ solution and its ultraspinning
cousin are deduced. Similar to the ultraspinning Kerr-Sen-AdS$_4$ black hole case, we
demonstrate that the ultraspinning dyonic Kerr-Sen-AdS$_4$ black hole does not always
violate the reverse isoperimetric inequality (RII) since the value of the isoperimetric
ratio can either be larger/smaller than, or equal to unity, depending upon the range of
the solution parameters, as is the case only with an electric charge. This property is
apparently distinct from that of the superentropic dyonic Kerr-Newman-AdS$_4$ black hole,
which always strictly violates the RII, although both of them have some similar properties
in other aspects, such as the horizon geometry and conformal boundary.
\end{abstract}

\maketitle

\section{Introduction}

Recently, a new class of ultraspinning AdS black holes \cite{PRL115-031101,JHEP0114127,
PRD89-084007}, in which one of their rotation angular velocities is boosted to the speed of
light, has attracted a lot of attention. This class of black hole has a finite horizon area
but with a noncompact horizon topology because there are two punctures at the north and south
poles of its spherical horizon. Interestingly, the ultraspinning black holes violate the RII
\cite{PRD84-024037,PRD87-104017}, and this means that the Schwarzschild-AdS black hole has
the maximum upper entropy. Because the ultraspinning black hole can exceed the maximum
entropy bound, it is therefore often called a ``superentropic" black hole. Moreover, it
is pointed out \cite{PRL115-031101} that one can obtain the corresponding superentropic
black hole solution by taking a simple ultraspinning limit from its usual rotating AdS
black hole. Such a solution generating trick is very simple: first, rewrite the metric of
the rotating AdS black hole in the rotating frame at infinity, then boost one of the rotation
angular velocities to the speed of light, and finally compactify the corresponding azimuthal
direction. Since then, a dozen of new superentropic black hole solutions \cite{JHEP0615096,
JHEP0816148,PRD95-046002,1702.03448,JHEP0118042} have been constructed from the known rotating
AdS black holes. Very recently, it has been suggested that the superentropic black hole can
also be obtained by running a conical deficit through the usual rotating AdS black hole
\cite{JHEP0220195}. On the other hand, various aspects of the superentropic black holes,
including thermodynamic properties \cite{PRL115-031101,JHEP0615096,PRD95-046002,1702.03448,
JHEP0118042,MPLA35-2050098,PRD101-086006,PRD101-024057,PLB807-135529}, horizon geometry
\cite{PRD89-084007,JHEP0615096,PRD95-046002}, Kerr/CFT correspondence \cite{PRD95-046002,
1702.03448,JHEP0816148}, and geodesic motion \cite{1912.03974}, etc, have also been extensively
studied.

Quite recently, we have studied some interesting properties of the Kerr-Sen-AdS$_4$ black holes
and their ultraspinning cousin in the four-dimensional gauged Einstein-Maxwell-Dilaton-Axion
(EMDA) theory \cite{PRD102-044007}. However, the black hole solution studied there only carries
an electric charge and is just a special and relatively simple case of the four-dimensional
gauged EMDA theory. It is then natural for us to extend that work to the more general dyonic
case, which serves as our motivation of the present work. First, we shall present the dyonic
generalization of the Kerr-Sen black hole solution and then include a nonzero negative
cosmological constant into it to obtain a dyonic Kerr-Sen-AdS$_4$ black hole. After that,
we will turn to investigate its ultraspinning counterpart. In the meanwhile, we will mainly
study their thermodynamical properties and verify that all the thermodynamical quantities
obtained for them perfectly obey both the extended law and the Bekenstein-Smarr mass formulas.

The organization of this article is outlined as follows. In Sec. \ref{II}, we first give
a brief introduction of the four-dimensional ungauged and gauged EMDA theories and summarize
the current already-known exact rotating charged black hole solutions in these supergravity
theories. In Sec. \ref{III}, we present the dyonic Kerr-Sen black hole solution and its
AdS$_4$ extension, and then turn to explore its thermodynamics. In Sec. \ref{IV}, with
the ultraspinning dyonic Kerr-Sen-AdS$_4$ black hole solution in hand, its thermodynamical
properties, horizon topology and conformal boundary, and the RII, etc, are subsequently
discussed. To this end, we derive new Christodoulou-Ruffini-like squared-mass formulae for
the dyonic Kerr-Sen-AdS$_4$ black hole and its ultraspinning cousin. By differentiating
them with respect to their individual thermodynamical variable, we get the expected
thermodynamical quantities which obey both the first law and the Bekenstein-Smarr mass
formulas without employing the chirality condition ($J = Ml$). After that, we impose the
chirality condition and derive the reduced form of the mass formulas. Finally, we show
that this ultraspinning dyonic Kerr-Sen-AdS$_4$ black hole does not always obey the RII,
since the value of the isoperimetric ratio can either be larger/smaller than, or equal to
unity, depending upon where the solution parameters lie in the parameters space. This property
is very similar to that of the ultraspinning Kerr-Sen-AdS$_4$ black hole \cite{PRD102-044007},
however, it signals a remarkable difference from the superentropic dyonic Kerr-Newman-AdS$_4$
black hole. Finally, the paper is ended up with our summaries in Sec. \ref{V}.

\section{EMDA supergravity theories and its already-knownrotating charged solutions}\label{II}

\subsection{Brief introduction to ungauged and gauged EMDA supergravity theories}

In 1992, Sen \cite{PRL69-1006} presented a stationary and axially symmetric solution
that describes a four-dimensional black hole beyond the Einstein-Maxwell theory. It
is the \emph{first} exact rotating charged solution in the low energy effective field
theory for the heterotic string theory. The bosonic sector of the four-dimensional
low-energy heterotic string theory contains the metric field $g_{\mu\nu}$, the $U(1)$
Abelian gauge field $A_{\mu}$, the dilaton scalar field $\phi$, and the three-order
totally antisymmetric tensor field $H_{\mu\nu\rho}$. Its Lagrangian reads
\be
\check{\mathcal{L}} = \sqrt{-g}\Big[R - \frac{1}{2}(\p\phi)^2
 -e^{-\phi}F^2 -\frac{1}{12}e^{-2\phi}H^2 \Big] \, ,
\ee
where $R$ is the Ricci scalar, $F_{\mu\nu}$ is the Faraday-Maxwell electromagnetic tensor
defined by $F = dA $, $F^2 = F_{\mu\nu}F^{\mu\nu}$, $(\p\phi)^2 = (\p_{\mu}\phi)(\p^{\mu}\phi)$,
and $H^2 = H_{\mu\nu\rho}H^{\mu\nu\rho}$.

In order to construct exact black hole solutions with a nonzero cosmological constant, the
three-form field must be dualized to an axion pseudoscalar field $\chi$ via the relation:
$H \equiv d\mathcal{B} -A\wedge\, F/4 = -e^{2\phi}\,{\star}d\chi$, where $\mathcal{B}$ is
an anti-symmetric two-form potential and the star operator represents the Hodge duality.
Then the resulted theory is also known as the EMDA supergravity theory, and accordingly,
the above Lagrangian can be rewritten in a different but completely equivalent form:
\bea
\hat{\mathcal{L}} &=& \sqrt{-g}\Big[R -\frac{1}{2}(\p\phi)^2 -\frac{1}{2}e^{2\phi}(\p\chi)^2
 -e^{-\phi}F^2\Big]  \nn \\
&&+\frac{\chi}{2}\epsilon^{\mu\nu\rho\lambda}F_{\mu\nu}F_{\rho\lambda} \, ,
\eea
where $\epsilon^{\mu\nu\rho\lambda}$ is the four-dimensional Levi-Civita antisymmetric tensor
density. From the equation of motion derived for the Abelian gauge potential $A$, one can define
its dual potential $B$ by: $e^{-\phi}\,{\star}F +\chi\, F = -dB$. It is very convenient to use
them to compute the electrostatic and magnetostatic potentials.

In the gauged version corresponding to the above EMDA theory, the corresponding Lagrangian
has the following form:
\be\label{gemda}
\mathcal{L} = \hat{\mathcal{L}} +\mathcal{V}(\phi,\chi)
 = \hat{\mathcal{L}} +\sqrt{-g}\big[4 +e^{-\phi} +e^{\phi}\big(1 +\chi^2\big)\big]/l^2 \, ,
\ee
with $l$ being the cosmological scale or the reciprocal of the gauge coupling constant.
Since the above Lagrangian is supplemented by a potential term related to the dilaton and
axion scalar fields, it is no longer possible to reexpress this Lagrangian into a dualized
version in terms of the three-form field that appeared in the ungauged version again.

The origin of the kinetic terms of matter fields and the scalar potential in the Lagrangian
(\ref{gemda}) of four-dimensional EMDA gauged supergravity can be attributed to the $S^7$
reduction of eleven-dimensional supergravity, which gives rise to SO(8) gauged $\mathcal{N}
= 8$ supergravity in four dimensions. A successive consistent truncation leads to $\mathcal{N}
= 2$, D = 4 U(1) Fayet-Iliopoulos gauged supergravity that admits a prepotential formalism
\cite{JGP23-111}, of which the well-known STU model \cite{NPB558-96,NPB554-237} endows with
the prepotential $\mathcal{F} = -X^1X^2X^3/X^0$. Further completion of the U(1)$^2$ truncation
of this STU model (by setting the charge parameters to be pair-wise equal), one arrives at
the SU(1,1)/U(1) model whose prepotential is given by $\mathcal{F} = -iX^0X^1$, sometimes
also known as ``$-iX^0X^1$ supergravity" \cite{PRD90-025029}. Finally, the D = 4 EMDA gauged
supergravity is obtained by setting one of the gauge field (or charge parameter) to zero.
See Ref. \cite{PRD90-025029} for a relatively recent review, and particularly Fig. 1 therein
that provides a schematic diagram to sketch the consistent reductions.

\subsection{Present status of stationary and axially symmetric solutions to the theories}

It is well-known that the most general stationary and axially symmetric class of type D
solution of the four-dimensional Einstein-Maxwell equations with an aligned electromagnetic
field is given by the 7-parameter family of the Pleba\'{n}ski-Demia\'{n}ski solution
\cite{AP98-98}, which contains the mass, NUT charge (dual mass), electric and magnetic
charges, rotation and acceleration parameters, and a nonzero cosmological constant. Until
now, no analogous extension of this general 7-parameter Pleba\'{n}ski-Demia\'{n}ski solution
has been found yet beyond the Einstein-Maxwell framework, including the gauged EMDA theory
that we are interested in here.

Soon after Sen \cite{PRL69-1006} gave the \emph{first} rotating charged black hole solution
in the above EMDA theory, a lot of intention has been paid to including more parameters into
it. In the ungauged EMDA theory, the existing methods to generalize the nonextremal Kerr-Sen
solution can be roughly classified into three categories:
\begin{enumerate}
\item[I.]
The brute force solving approach. As a representative of this method, the work \cite{PRL74-1276}
adopted a suitable ansatz for the line element but a very restrictive one for the U(1) gauge
potential, namely, the non-null electromagnetic field is not the most general aligned one.
The solution presented in Ref. \cite{PRL74-1276} can be thought of as a very special dyonic
generalization of the Kerr-Sen solution.

\item[II.]
Three different solution generating methods that correspond to different coset (potential)
spaces depending upon dimensional reduction from distinct superstring and supergravity
theories as follows:

II a) The Hassan-Sen method \cite{NPB375-103}. The Kerr-Sen solution was initially generated
\cite{PRL69-1006} from the famous Kerr solution via this approach by using a $9\times 9$ coset
matrix. Later, the same method was subsequently applied in Refs. \cite{CQG14-999,PLB782-594}
to generate accelerating, rotating, charged black holes in the low energy heterotic string
theory, respectively, from some type D (Pleba\'{n}ski-Demia\'{n}ski) metrics and its special
case, namely, the accelerating Kerr solution. However, it should be pointed out here that it
is a very difficult or rather challenging task to further include a nonzero cosmological
constant into the so-obtained accelerating, rotating, charged solution.

II b) The Sp(4,R)/U(2) potential space method \cite{PRL74-2863,JMP36-5023}. Starting from
the Kerr-NUT solution, a new rotating dyonic black hole solution was generated in Ref.
\cite{PRD50-7394} by using the symmetry of this potential space. Its solution generating
process is completely equivalent to the Hassan-Sen method when employed to the Kerr-NUT
(or only Kerr) solution firstly, and then followed by implementing a necessary gauge
transformation and the generalized electromagnetic duality transformation \cite{PRD50-7394}
to the obtained gauge potential. After being reexpressed in terms of observable physical
quantities, namely, the mass, NUT charge, electric and magnetic charge as well as rotation
parameters, the resulted solution can be thought of as a dyonic NUT extension of the Kerr-Sen
solution, which shall be named as the dyonic Kerr-Sen-NUT solution hereafter. However, the
gauge potential had not been explicitly given in Ref. \cite{PRD50-7394}, and its spatial
component still needs to be worked out via the dual of the magnetic scalar potential.

II c) Subgroup of larger coset spaces: O(4,4)/O(1,1)$^4$ \cite{NPB717-246} and SO(4,4)/SL(2,R)$^4$
\cite{CQG31-022001,PRD90-025029}. The coset matrix representation of the EMDA theory can be
viewed as a subset of these more complicated coset matrix representations of the four-dimensional
STU supergravity theory, and the dyonic Kerr-Sen-NUT solution \cite{PRD50-7394} is just a special
case of the pair-wise equal charge parameters that was introduced in Refs. \cite{NPB717-246,
PRD90-025029}. The charging parameters used in \cite{NPB717-246} are two electric and two
magnetic ones, and the seed metric adopted in the main context of that paper is the Kerr
solution, but in the appendix, the authors had already considered the Kerr-NUT solution and
the most general Pleba\'{n}ski-Demia\'{n}ski metric as the seed solution too. In particular,
they had presented the explicit expressions for the obtained solutions in the case of the
pair-wise equal charge parameters, and mentioned that they failed to make a generalization
so as to include a nonzero cosmological constant when a nonvanishing acceleration parameter
is turned on. On the other hand, the charging parameters chosen in \cite{PRD90-025029} are
enlarged to four electric and four magnetic ones, but their seed metric is only the Kerr-NUT
solution. Incidentally, it should be pointed out that the accelerating Kerr-Sen and accelerating
Kerr-NUT-Sen solution generated in Ref. \cite{PLB782-594} are just special cases of those were
previously obtained in Ref. \cite{NPB717-246} to possess the pair-wise equal charge parameters,
although they are re-derived by using a simpler Hassan-Sen method.

\item[III.]
The Belinsky-Zakharov inverse scattering technique \cite{JETP48-985} was extended in
\cite{PRD65-024024} to act on the Sp(4,R)/U(2) coset space to get the dyonic Kerr-Sen-NUT
solution that had already been obtained in Ref. \cite{PRD50-7394}.
\end{enumerate}

In the case of the four-dimensional gauged STU supergravity theory, no solution generating
technique can be used to get the rotating charged AdS solution, of which some subclasses
had been already constructed \cite{NPB717-246,PRD89-065003} only in the special case of
the pair-wise equal charge parameters\footnote{Rotating charged AdS solutions with one
parameter less or more than those of Ref. \cite{PRD89-065003} were successively constructed
in \cite{JHEP0114127,JHEP0916088}. It is declared in \cite{JHEP0916088} that the extra
parameter $\beta$ represents a scalar hair, but it is unclear to us whether or not it is a
redundant one.}, and in the consistent truncation cases with single-charge \cite{CQG28-032001,
PRD83-121502} and two-charge \cite{CQG28-175004}, respectively. However, they are not written
in the simple or concise expressions, but expressed in the complicated forms in terms of
charging parameters \cite{NPB717-246,PRD89-065003}. The most general rotating charged AdS$_4$
solution with generic unequal values of four pairs of electromagnetic charging parameters that
generalizes the generic Chow-Comp\`{e}re solution \cite{PRD90-025029} still remain elusive
till now, let alone further introducing a nonzero acceleration parameter. The latter generic
solution to four-dimensional gauged STU supergravity theory should contain 13 parameters that
represent the mass, NUT charge, rotation and acceleration, four electric and four magnetic
charge parameters, and a nonzero cosmological constant.

\section{Dyonic Kerr-Sen black hole and its AdS$_4$ extension}\label{III}

\subsection{A new simple form of dyonic Kerr-Sen solution}

Although the dyonic NUT generalization of the Kerr-Sen black hole solution was already
given in Ref. \cite{PRD50-7394} sixteen years ago, we feel that its expressions for the
solution is not suitable to our aim here. In particular, the angular component for the
U(1) gauge potential was not explicitly given there. The solution contains a minimal set
of five physical quantities, which correspond to the mass, NUT charge, electric and magnetic
charges as well as rotation parameter, respectively, whilst the dilaton scalar and axion
pseudoscalar charges are not independent parameters but related to these five parameters
mentioned above. In this paper, we will consider a slightly simpler case without the NUT
charge, namely, the dyonic extension of the Kerr-Sen solution. In other words, we shall
extend our previous work \cite{PRD102-044007} to a more general case by adding only a
nonzero magnetic charge to the Kerr-Sen black hole.

Written in terms of the Boyer-Lindquist coordinates, the line element, the Abelian gauge
potential and its dual, as well as the dilaton scalar and axion pseudoscalar fields of
the dyonic Kerr-Sen black hole are given in the following exquisite forms:
\be\label{KSdy}
\begin{aligned}
 d\hat{s}^2 &= -\frac{\hat{\Delta}(r)}{\hat{\Sigma}}\hat{X}^2
 +\frac{\hat{\Sigma}}{\hat{\Delta}(r)}dr^2 +\hat{\Sigma} d\theta^2
  +\frac{\sin^2\theta}{\hat{\Sigma}}\hat{Y}^2 \, , \\
\hat{A} &= \frac{q\big(r -p^2/m\big)}{\hat{\Sigma}}\hat{X}
 -\frac{p\cos\theta}{\hat{\Sigma}}\hat{Y} \, , \\
\hat{B} &= \frac{p\big(r -p^2/m\big)}{\hat{\Sigma}}\hat{X}
 +\frac{q\cos\theta}{\hat{\Sigma}}\hat{Y} \, , \\
 e^{\hat{\phi}} &= \frac{r^2 +(a\cos\theta +k)^2}{\hat{\Sigma}} \, , \quad
 \hat{\chi} = 2\frac{k\, r -d(a\cos\theta +k)}{r^2 +(a\cos\theta +k)^2} \, ,
\end{aligned}
\ee
where
\bea
&&\hat{X} = dt -a\sin^2\theta\, d\hat{\varphi} \, , \quad
 \hat{Y} = a\, dt -\big(r^2 -2dr -k^2 +a^2\big)d\hat{\varphi} \, , \nn \\
&&\hat{\Delta}(r) = r^2 -2dr -2m(r -d) -k^2 +a^2 +p^2 +q^2 \, , \nn \\
&&\quad~ \hat{\Sigma} = r^2 -2dr -k^2 +a^2\cos^2\theta \, ,  \nn
\eea
in which, $d = (p^2 -q^2)/(2m)$ and $k = pq/m$ represent the dilaton scalar and axion
pseudoscalar charges, and the mass, electric and magnetic charges as well as angular
momentum of the black hole are: $M = m$, $Q = q$, $P = p$, and $J = ma$, respectively.
When the magnetic charge vanishes ($p = 0$), the axion charge vanishes too ($k = 0$),
and then the solution reduces to the Kerr-Sen case previously considered in our previous
work \cite{PRD102-044007} with $b = -d = q^2/(2m)$. On the other hand, in the special
case when $p = q$, the dilaton charge will completely vanish ($d = 0$). Note that there
is a useful quadratic relation: $d^2 +k^2 = \big(p^2 +q^2\big)^2/(4m^2)$. In addition,
if wishes, one can work with another radial coordinate by shifting $r\to r +p^2/m$, or
$r\to r +d$ to make the expressions more symmetric about $(p, q, d, k)$ (especially in
the case when the NUT parameter is turned on).

It should be emphasized that in the above, we have chosen a concrete gauge choice so that
the temporal components of both Abelian gauge potentials completely vanish at infinity, in
the meanwhile, their angular components simultaneously become $p\cos\theta$ and $-q\cos\theta$
there, respectively. Our arguments for this gauge choice go as follows.

Generally speaking, the Abelian gauge potential $\hat{A}$ at the infinity has the asymptotic
form: $\hat{A}_{\infty} = \Phi_{\infty}dt +p(\cos\theta \pm C)d\hat{\varphi}$. In the standard
monopole gauge theory, the constant $C$ may usually take three different values: $C = 0$ on
the equator, and $C = \pm 1$ at the north and south poles ($\theta = 0, \pi$) to eliminate
the Dirac string singularities at two poles, respectively. In the static case, the constant
$C$ can take any one of these three values, and this does not lead to any serious problem.
But in the rotating case, if the constant $C$ takes the above gauge choice at infinity, namely,
$C = 0, \pm 1$ on the equator and at the north and south poles, respectively, then it would
give rise to an odd result for the expressions of the electrostatic potential on the horizon:
they are not identical to each other, for instance, on the equator, at two poles and elsewhere.
This obviously contradicts the common sense that the electrostatic potential should be a unique
constant everywhere on the event horizon, and it means that the electric charge is not uniformly
distributed on the horizon. To ensure that the electrostatic potential in the rotating case is
equal everywhere on the horizon, one can only set $C = 0$ so that $\hat{A}_\phi = p\cos\theta$
at infinity. On the other hand, any choice of $\Phi_{\infty}$ does not change the difference of
the electrostatic potential on the event horizon and that measured by an observer at infinity:
$\Phi = \Phi_+ -\Phi_{\infty}$. Frequently, two conventional options for the temporal component
of the gauge potential $\hat{A}$ are either $\Phi_{\infty} = 0$ or $\Phi_+ = 0$ (equivalently,
$\Phi_{\infty} = -\Phi_+$). Similar discussions completely apply to the dual Abelian gauge
potential $\hat{B}$ too. Here we would like to work with the temporal gauge choice $\Phi_{\infty}
= \Psi_{\infty} = 0$. However, it should be noted that different choices of temporal gauge of
the U(1) potentials correspond to different thermodynamical ensembles and therefore different
thermodynamical grand potentials. One can have a total of four possibilities by fixing any one
combination of the charges, electrostatic and magnetostatic potentials: ($Q, P$), ($Q, \Psi$),
($P, \Phi$) and ($\Phi, \Psi$). Taking into account of these, our discussions below about
thermodynamics in the dyonic case should correspond to the thermodynamical ensemble with
the static potentials ($\Phi, \Psi$) fixed.

\subsection{Dyonic Kerr-Sen-AdS$_4$ solution}

We now add a nonzero negative cosmological constant into the above dyonic Kerr-Sen black hole
solution and obtain an exact AdS$_4$ black hole solution to the gauged EMDA theory. Expressed
in terms of the Boyer-Lindquist coordinates with the frame rotating at infinity, the dyonic
Kerr-Sen-AdS$_4$ black hole solution can be written as follows:
\be\label{KSAdSdy}
\begin{aligned}
d\bar{s}^2 &= -\frac{\bar{\Delta}_r}{\bar{\Sigma}}\bar{X}^2
 +\frac{\bar{\Sigma}}{\bar{\Delta}_r}dr^2 +\frac{\bar{\Sigma}}{\bar{\Delta}_\theta}d\theta^2
  +\frac{\bar{\Delta}_\theta\sin^2\theta}{\bar{\Sigma}}\bar{Y} \, , \\
\bar{A} &= \frac{q\big(r -p^2/m\big)}{\bar{\Sigma}}\bar{X}
 -\frac{p\cos\theta}{\bar{\Sigma}}\bar{Y} \, , \\
\bar{B} &= \frac{p\big(r -p^2/m\big)}{\bar{\Sigma}}\bar{X}
 +\frac{q\cos\theta}{\bar{\Sigma}}\bar{Y} \, , \\
e^{\bar{\phi}} &= \frac{r^2 +(a\cos\theta +k)^2}{\bar{\Sigma}} \, , \quad
\bar{\chi} = 2\frac{k\, r -d(a\cos\theta +k)}{r^2 +(a\cos\theta +k)^2} \, ,
\end{aligned}
\ee
where a bar is designed to distinct the present expressions from their corresponding ones in
the ungauged case. Note that $\bar{\Sigma} = \hat{\Sigma} = r^2 -2dr -k^2 +a^2\cos^2\theta$
as before, but now we have
\bea
&&\bar{X} = dt -\frac{a\sin^2\theta}{\Xi}d\bar{\varphi} \, , \quad
\bar{Y} = a\, dt -\frac{r^2 -2dr -k^2 +a^2}{\Xi}d\bar{\varphi} \, , \nn \\
&&\bar{\Delta}_r = \bigg(1 +\frac{r^2 -2dr -k^2}{l^2}\bigg)\big(r^2 -2dr -k^2 +a^2\big)\nn \\
&&\qquad\quad -2m(r -d) +p^2 +q^2 \, , \nn \\
&&\bar{\Delta}_\theta = 1 -\frac{a^2}{l^2}\cos^2\theta \, , \qquad
 \Xi = 1 -\frac{a^2}{l^2} \, . \nn
\eea
Clearly, the above solution (\ref{KSAdSdy}) simply reduces to the dyonic Kerr-Sen black hole
solution (\ref{KSdy}) when the cosmological constant vanishes.

Since the gauged EMDA theory is a successive consistent truncation of the four-dimensional
gauged STU supergravity, therefore, similar to the purely electric-charged case as mentioned
in our previous article \cite{PRD102-044007}, the above AdS$_4$ black hole solution can be
thought of as a special case of those obtained in Refs. \cite{NPB717-246,PRD89-065003}, where
more general solutions with the pair-wise equal charge parameters have been constructed in
the gauged ``$-iX^0X^1$ supergravity" model. However, the solution presented here is slightly
simpler than those given there especially by the radial structure function, and is more
convenient for further investigations.

\subsection{Thermodynamics}

Now we would like to explore thermodynamics of the dyonic Kerr-Sen-AdS$_4$ black hole
(\ref{KSAdSdy}). One can compute all associated thermodynamic quantities via the standard
method and express them as follows:
\bea\label{Therm}
&&\bar{M} = \frac{m}{\Xi} \, , \qquad \bar{J} = \frac{ma}{\Xi^2} \, , \qquad
\bar{Q} = \frac{q}{\Xi} \, , \qquad \bar{P} = \frac{p}{\Xi} \, , \nn \\
&&\bar{S} = \frac{\pi}{\Xi}\big(r_+^2 -2dr_+ -k^2 +a^2\big) \, , \quad
\bar{\Omega} = \frac{a\Xi}{r_+^2 -2dr_+ -k^2 +a^2} \, , \quad \nn  \\
&&\bar{\Phi} = \frac{q\big(r_+ -p^2/m\big)}{r_+^2 -2dr_+ -k^2 +a^2} \, , \quad
\bar{\Psi} = \frac{p\big(r_+ -p^2/m\big)}{r_+^2 -2dr_+ -k^2 +a^2} \, , \nn  \\
&&\bar{T} = \frac{\bar{\Delta}^{\prime}_{r_+}}{4\pi\big(r_+^2 -2dr_+ -k^2 +a^2\big)} \nn \\
&&\quad = \frac{(r_+ -d)\big(2r_+^2 -4dr_+ -2k^2 +a^2 +l^2\big) -ml^2}{2\pi
 \big(r_+^2 -2dr_+ -k^2 +a^2\big)l^2} \, ,
\eea
where the location of the event horizon $r_+$ is the largest root of equation:
$\bar{\Delta}_{r_+} = 0$.

In the above, we have only listed the final expressions for these thermodynamic quantities
while omitting the computing process. In particular, we have adopted the conformal completion
method provided in Ref. \cite{PRD73-104036} (especially its subsection IV. A is very useful
and relevant to our aim) to compute the conserved charges (mass and angular momentum) given
in Eqs. (\ref{Therm}) and (\ref{MOV}) in the rotating and rest frames at infinity, respectively.
On the other hand, due to good fall-off behaviors of the vector and scalar fields, the electric
and magnetic charges can be computed by the Gauss-type law integral:
\be
\begin{aligned}
Q &= \frac{1}{4\pi}\int\star\, d\bar{A} = \frac{1}{4\pi}\int d\bar{B} \, , \\
P &= \frac{1}{4\pi}\int\star\, d\bar{B} = \frac{-1}{4\pi}\int d\bar{A} \, .
\end{aligned}
\ee
In addition, by virtue of the specific gauge choice argued in the last section, the
electrostatic and magnetostatic potentials are simply given by
\be
\bar{\Phi} = (A_{\mu}\bar{\chi}^{\mu})|_{r=r_+} \, , \quad
\bar{\Psi} = (B_{\mu}\bar{\chi}^{\mu})|_{r=r_+} \, ,
\ee
where $\bar{\chi} = \p_t +\bar{\Omega}\p_{\bar{\varphi}}$ is the Killing vector normal to the
event horizon. Similarly, one can compute their corresponding thermodynamic expressions in the
rest frame at infinity also.

It is not difficult to check that the thermodynamic quantities (\ref{Therm}) obey the
Bekenstein-Smarr mass formulas
\be
\bar{M} = 2\bar{T}\bar{S} +2\bar{\Omega}\bar{J} +\bar{\Phi}\bar{Q}
 +\bar{\Psi}\bar{P} -2\bar{V}\bar{\mathcal{P}} \, ,
\ee
where $\bar{V}$ is the thermodynamic volume
\be
\bar{V} = \frac{4}{3}(r_+ -d)\bar{S}
 = \frac{4\pi}{3\Xi}(r_+ -d)\big(r_+^2 -2dr_+ -k^2 +a^2\big) \, ,
\ee
which is conjugate to the pressure $\bar{\mathcal{P}} = 3/(8\pi\,l^2)$, and can be alternatively
evaluated via the method put forward in Ref. \cite{PRD84-024037} (See III. B therein)
\be
\bar{V} = \frac{l^2}{3}\int_d^{r_+}dr\int_0^{2\pi}d\bar{\varphi}\int_0^{\pi}d\theta\,
 \sqrt{-\bar{g}}\mathcal{V}(\bar{\phi},\bar{\chi}) \, ,
\ee
where the lower limit of $r$-integration must be taken to be $r = d$. However, the first law
becomes a differential identity only
\be
d\bar{M} = \bar{T}d\bar{S} +\bar{\Omega}d\bar{J} +\bar{\Phi}d\bar{Q} +\bar{\Psi}d\bar{P}
 +\bar{V}d\bar{\mathcal{P}} +\bar{J}d\Xi/(2a) \, .
\ee
The reason for this is that we have worked with a frame rotating at infinity.

One can transform the above dyonic Kerr-Sen-AdS$_4$ solution into the frame rest at infinity
via a simple coordinate transformation: $\bar{\varphi}\rightarrow \tilde{\varphi} -al^{-2}t$.
After a cumbersome computation (using the method mentioned above) for those thermodynamic
quantities in this rest frame, it is easy to observe that only the mass, the angular velocity
and the thermodynamic volume are different from those given in Eq. (\ref{Therm}) and related
by the following expressions:
\be\label{MOV}
\tilde{M} = \bar{M} +\frac{a}{l^2}\bar{J} = \frac{m}{\Xi^2} \, , \quad
 \tilde{\Omega} = \bar{\Omega} +\frac{a}{l^2} \, , \quad
\tilde{V} = \bar{V} +\frac{4\pi}{3}a\bar{J} \, .
\ee

Now it is easy to verify that thermodynamic quantities can indeed fulfil both the standard
forms of the first law and the Bekenstein-Smarr mass formula simultaneously:
\be
\begin{aligned}
d\tilde{M} &= \bar{T}d\bar{S} +\tilde{\Omega}d\bar{J}
 +\bar{\Phi}d\bar{Q} +\bar{\Psi}d\bar{P} +\tilde{V}d\bar{\mathcal{P}} \, , \\
\tilde{M} &= 2\bar{T}\bar{S} +2\tilde{\Omega}\bar{J}
 +\bar{\Phi}\bar{Q} +\bar{\Psi}\bar{P} -2\tilde{V}\bar{\mathcal{P}} \, .
\end{aligned}
\ee
In addition, one can show that the above differential and integral mass formulae can be
derived from the following Christodoulou-Ruffini-like squared-mass formulas
\be
\tilde{M}^2 = \bigg(1 +\frac{8\bar{\mathcal{P}}\bar{S}}{3}\bigg)\bigg[\bigg(1
 +\frac{8\bar{\mathcal{P}}\bar{S}}{3}\bigg)\frac{\bar{S}}{4\pi}
  +\frac{\pi\,\bar{J}^2}{S} +\frac{\bar{P}^2 +\bar{Q}^2}{2}\bigg] \, . \quad
\ee
When the magnetic charge $P$ vanishes, all the above thermodynamic formulae can consistently
reduce to those obtained in Ref. \cite{PRD102-044007} for the purely electric-charged
Kerr-Sen-AdS$_4$ case.

\section{Ultraspinning dyonic Kerr-Sen-AdS$_4$ black hole}\label{IV}

\subsection{The ultraspinning dyonic solution}

Following \cite{PRD102-044007}, we redefine $\varphi = \bar{\varphi}/\Xi$ and then just need
to set $a = l$ in the above dyonic Kerr-Sen-AdS$_4$ black hole solution (\ref{KSAdSdy}) to
construct its corresponding ultraspinning version as follows:
\be\label{SEKSAdSdy}
\begin{aligned}
ds^2 &= -\frac{\Delta(r)}{\Sigma}X^2 +\frac{\Sigma}{\Delta(r)} dr^2
 +\frac{\Sigma}{\sin^2\theta} d\theta^2 +\frac{\sin^4\theta}{\Sigma}Y^2 \, , \\
A &= \frac{q\big(r -p^2/m\big)}{\Sigma}X -\frac{p\cos\theta}{\Sigma}Y \, ,  \\
B &= \frac{p\big(r -p^2/m\big)}{\Sigma}X +\frac{q\cos\theta}{\Sigma}Y \, , \\
 e^{\phi} &= \frac{r^2 +(l\cos\theta +k)^2}{\Sigma} \, , \quad
 \chi = 2\frac{k\, r -d(l\cos\theta +k)}{r^2 +(l\cos\theta +k)^2} \, ,
\end{aligned}
\ee
where
\bea
&& X = dt -l\sin^2\theta\, d\varphi \, , \quad
Y = l\, dt -\big(r^2 -2dr -k^2 +l^2\big)d\varphi \, , \nn \\
&& \Delta(r) = \big(r^2 -2dr -k^2 +l^2\big)^2l^{-2} -2m(r -d) +p^2 +q^2 \nn \\
&&\qquad = \big[\big(r+q^2/m\big)\big(r-p^2/m\big) +l^2\big]^2l^{-2}
 -2m\big(r-p^2/m\big) \, , \nn \\
&&\quad~\Sigma = r^2 -2dr -k^2 +l^2\cos^2\theta \nn \\
&&\qquad = \big(r+q^2/m\big)\big(r-p^2/m\big) +l^2\cos^2\theta \, . \nn
\eea
Note that in the above ultraspinning dyonic solution, the period of $\varphi$ is now assumed
to take a dimensionless parameter $\mu$ rather than $2\pi$.

With an exact solution of the ultraspinning dyonic Kerr-Sen-AdS$_4$ black hole in hand, the
remaining main task of this work is  to study its various interesting basic properties, such
as its thermodynamical properties, the horizon topology and conformal boundary, and the RII,
etc.

\subsection{Various mass formulae}

First, let us focus on thermodynamics of the ultraspinning dyonic Kerr-Sen-AdS$_4$ black hole.
As before, one can obtain the following expressions of its fundamental thermodynamic quantities
through the standard method:
\bea\label{therm}
&& M = \frac{\mu}{2\pi}m \, , \quad J = \frac{\mu}{2\pi}ml = Ml \, , \quad
Q = \frac{\mu}{2\pi}q \, , \quad P = \frac{\mu}{2\pi}p \, , \nn \\
&& S = \frac{\mu}{2}\big(r_+^2 -2dr_+ -k^2 +l^2\big) \, , \quad
\Omega = \frac{l}{r_+^2 -2dr_+ -k^2 +l^2} \, , \nn \\
&& \Phi = \frac{q\big(r_+ -p^2/m\big)}{r_+^2 -2dr_+ -k^2 +l^2} \, , \quad
\Psi = \frac{p\big(r_+ -p^2/m\big)}{r_+^2 -2dr_+ -k^2 +l^2} \, , \nn \\
&& T  = \frac{\Delta^{\prime}(r_+)}{4\pi\big(r_+^2 -2dr_+ -k^2 +l^2\big)} \nn \\
&&\quad = \frac{r_+ -d}{\pi\, l^2} -\frac{m}{2\pi\big(r_+^2 -2dr_+ -k^2 +l^2\big)} \, ,
\eea
in which the location of the event horizon $r_+$ is now determined by the largest root of
equation: $\Delta(r_+) = 0$.

As mentioned in the last section, one good way to compute the mass and angular momentum in
the ultraspinning case is to adopt the conformal completion technique that is explicitly
elucidated in Ref. \cite{PRD73-104036} since the line element of the conformal boundary is
not a diagonal metric. It is worthy to point out that all of the thermodynamic quantities
given in Eq. (\ref{therm}) are obtained by applying the same method as mentioned above.
Because a detailed discussion about the computation of all the thermodynamic quantities for
the ultraspinning AdS solution had already addressed in the subsection II. A of our previous
work \cite{PRD101-024057}, so we omit the computing process here.

Note that there is a chirality condition ($J = Ml$) that constrains the angular momentum
and the mass, and the angular velocity $\Omega$ is that of the event horizon because the
ultraspinning dyonic black hole is rotating at the speed of light at infinity.

Now it can be shown that the above thermodynamical quantities completely fulfil both the
first law and the Bekenstein-Smarr mass formulas:
\bea
dM &=& TdS +\Omega\, dJ +\Phi\, dQ +\Psi\, dP +V d\mathcal{P} +K d\mu \, , \label{FL} \\
M &=& 2TS +2\Omega\, J + \Phi\, Q +\Psi\, P -2V\mathcal{P} \label{Smarr} \, ,
\eea
in which the thermodynamic volume and a new chemical potential
\bea
V &=& \frac{4}{3}(r_+ -d)S = \frac{2}{3}\mu(r_+ -d)
 \big(r_+^2 -2dr_+ -k^2 +l^2\big)\, , \quad \label{Ve} \\
K &=& m\frac{l^2 -\big(r_+ +q^2/m\big)\big(r_+ -p^2/m\big)}{4\pi
 \big(r_+^2 -2dr_+ -k^2 +l^2\big)} \, , \label{K}
\eea
are conjugate to the pressure $\mathcal{P} = 3/(8\pi\, l^2)$ and the dimensionless parameter
$\mu$, respectively. Note that $V$ can be also computed as
\be
V = \frac{l^2}{3}\int_d^{r_+}dr\int_0^{\mu}d\varphi\int_0^{\pi}d\theta\,
 \sqrt{-g}\mathcal{V}(\phi,\chi) \, ,
\ee
where the upper limit of $\varphi$-integration is changed to $\varphi = \mu$, while the lower
limit of $r$-integration still is $r = d$. Presumably, the chemical potential $K$ might be
obtained alternatively in the context of an angular gravitational tension \cite{PRD86-081501}
associated to the rotational symmetry of the black hole or in term of the thermodynamic length
\cite{JHEP0517116} related to the conical defect.

In the subsection II. E of our previous work \cite{PRD101-024057} (and then III. C of Ref.
\cite{PRD102-044007}), we have made, for the first time, an attempt to relate the thermodynamic
quantities of the ultraspinning solution to those of its usual rotating AdS solution. This
is based upon such a naive belief that the ultraspinning solution is obtained by taking the
$a\to l$ limit, so were their thermodynamic quantities also. Similar to what was done in
Refs. \cite{PRD101-024057,PRD102-044007}, now we simply generalize to the dyonic case, and
proceed to assume the following simple relations
\bea\label{rel}
&&M = \frac{\mu\Xi\bar{M}}{2\pi} \, , \quad J = \frac{\mu\Xi^2\bar{J}}{2\pi} \, , \quad
Q = \frac{\mu\Xi\bar{Q}}{2\pi} \, , \quad P = \frac{\mu\Xi\bar{P}}{2\pi} \, ,  \quad \nn \\
&&\Omega = \frac{\bar{\Omega}}{\Xi} \, , \qquad S = \frac{\mu\Xi\bar{S}}{2\pi} \, , \qquad
 V = \frac{\mu\Xi\bar{V}}{2\pi} \, ,  \\
&&T = \bar{T} \, , \qquad \Phi = \bar{\Phi} \, , \qquad
\mathcal{P} = \bar{\mathcal{P}} \, , \nn
\eea
and take the ultraspinning limit: $a \to l$. Then we can find that the above thermodynamic
quantities presented in Eq. (\ref{therm}) for the ultraspinning dyonic Kerr-Sen-AdS$_4$
black hole can also be obtained straightforwardly from those of its corresponding usual
black hole in a frame rotating at infinity.

In Ref. \cite{PRD102-044007}, a new Christodoulou-Ruffini-like squared-mass formulas was
derived for the ultraspinning Kerr-Sen-AdS$_4$ black hole. Now we expect to generalize it
to the ultraspinning dyonic case. Rewriting the event horizon equation ($\Delta_{r_+} = 0$)
as
\be
\frac{S^2}{\pi\, l^2} +\pi\big(P^2 +Q^2\big) = \mu\, M(r_+ -d) \, ,
\ee
then exploiting $3/l^2 = 8\pi\,\mathcal{P}$, we can obtain: $r_+ = d +\big[8\mathcal{P}S^2
+3\pi\big(P^2 +Q^2\big)\big]/(3\mu\, M)$. Now, we substitute it into the entropy: $S = \mu
\big(r_+^2 -2dr_+ -k^2 +l^2\big)/2$ and use $d = \pi\big(P^2 -Q^2\big)/(\mu\, M)$ and $k =
2\pi\, PQ/(\mu\, M)$ as well as the chirality condition ($J = Ml$) to get a useful identity:
\be
M^2 = \frac{8\mathcal{P}S}{3\mu}\bigg[\frac{4\mathcal{P}}{3}S^2 +\pi\big(P^2 +Q^2\big)\bigg]
 +\frac{\mu\, J^2}{2S} \, ,
\label{sqm}
\ee
which is our expected Christodoulou-Ruffini-like squared-mass formulas for the ultraspinning
dyonic Kerr-Sen-AdS$_4$ black hole. We point out that this squared-mass formulas (\ref{sqm})
consistently reduces to the one obtained in the ultraspinning Kerr-Sen-AdS$_4$ black hole
case \cite{PRD102-044007} when the magnetic charge $P$ is turned off.

Supposing temporarily that there exists no chirality condition ($J = Ml$) at all, then it is
clear from Eq. (\ref{sqm}) that the thermodynamical quantities $S, J, Q, P, \mathcal{P}$ and
$\mu$ can be treated as independent thermodynamical variables and consist of an entire set of
extensive variables for the fundamental functional relation $M = M(S, J, Q, P, \mathcal{P},
\mu)$. In this way, as is done in Refs. \cite{PRD101-024057,PRD102-044007,PRD21-884,CQG17-399,
PLB608-251,CPL23-1096}, the differentiation of the above squared-mass formula (\ref{sqm})
with respect to one formal variable of the whole set ($S, J, Q, P, \mathcal{P}, \mu$) and
simultaneously fixing the remaining ones, respectively, lead to their corresponding conjugate
quantities as expected. Subsequently, one can obtain the differential first law (\ref{FL})
and the integral Bekenstein-Smarr relation (\ref{Smarr}) with the conjugate thermodynamic
potentials correctly reproduced by the ordinary Maxwell relations.

Let us now demonstrate the above conclusion in more detail. Differentiating the squared-mass
formula (\ref{sqm}) with respect to the entropy $S$ yields the conjugate Hawking temperature:
\bea
T &=& \frac{\p\, M}{\p\, S}\bigg|_{(J,Q,P,\mathcal{P},\mu)}
 = \frac{8\mathcal{P}}{3\mu\, M}\bigg[\frac{8\mathcal{P}}{3}S^2
  +\pi\big(P^2 +Q^2\big)\bigg] -\frac{M}{2S} \nn \\
&=& \frac{r_+ -d}{\pi\, l^2} -\frac{m}{2\pi\big(r_+^2 -2dr_+ -k^2 +l^2\big)} \, ,
\eea
and the corrected angular velocity, the electrostatic and magnetostatic potentials, which
are conjugate to $J$, $Q$, and $P$, respectively, can be computed as
\bea
\Omega &=& \frac{\p\, M}{\p\, J}\bigg|_{(S,Q,P,\mathcal{P},\mu)} = \frac{\mu\, J}{2MS} \nn \\
 &=& \frac{l}{r_+^2 -2dr_+ -k^2 +l^2} \, , \hspace*{2cm} \\
\Phi &=& \frac{\p\, M}{\p\, Q}\bigg|_{(S,J,P,\mathcal{P},\mu)}
 = \frac{8\pi\, \mathcal{P}Q}{3\mu\, M}S \nn \\
 &=& \frac{q\big(r_+ -p^2/m\big)}{r_+^2 -2dr_+ -k^2 +l^2} \, , \\
\Psi &=& \frac{\p\, M}{\p\, P}\bigg|_{(S,J,Q,\mathcal{P},\mu)}
 = \frac{8\pi\, \mathcal{P}P}{3\mu\, M}S \nn \\
 &=& \frac{p\big(r_+ -p^2/m\big)}{r_+^2 -2dr_+ -k^2 +l^2} \, .
\eea
Similarly, by the differentiation of the squared-mass formula (\ref{sqm}) with respect
to the pressure $\mathcal{P}$ and the dimensionless parameter $\mu$, one can obtain the
thermodynamical volume and a new chemical potential
\bea
V &=& \frac{\p\, M}{\p\mathcal{P}}\bigg|_{(S,J,Q,P,\mu)}
 = \frac{4S}{3\mu\, M}\Big[\frac{8\mathcal{P}}{3}S^2 +\pi\big(P^2 +Q^2\big)\Big] \nn \\
&=& \frac{4}{3}(r_+ -d)S \, , \quad \\
K &=& \frac{\p\, M}{\p\mu}\bigg|_{(S,J,Q,P,\mathcal{P})}
 = \frac{M}{2\mu} -\frac{8\mathcal{P}S}{3\mu^2 M}\Big[\frac{4\mathcal{P}}{3}S^2
 +\pi\big(P^2 +Q^2\big)\Big] \nn \\
&=& m\frac{l^2 -\big(r_+ +q^2/m\big)\big(r_+ -p^2/m\big)}{4\pi
 \big(r_+^2 -2dr_+ -k^2 +l^2\big)} \, .
\eea

All the above conjugate quantities correctly reproduce those expressions previously presented
in Eqs. (\ref{therm}), (\ref{Ve}) and (\ref{K}). Using all these thermodynamical conjugate pairs,
it is trivial to check that both the differential first law (\ref{FL}) and the integral mass
formula (\ref{Smarr}) are completely satisfied at the same time.

\subsection{Chirality condition and reduced mass formulae}

Now we would like to discuss in details about the impact of the chirality condition ($J = Ml$)
on the thermodynamical relations of the ultraspinning dyonic Kerr-Sen-AdS$_4$ black hole. By
virtue of the existence of the chirality condition ($J = Ml$), three thermodynamical quantities
($M, J, \mathcal{P}$) are, in fact, not truly independent of each other, and there is a
constraint relation among them
\be
J^2 = 3M^2/(8\pi\mathcal{P}) \, , \label{const}
\ee
which implies that the above ultraspinning dyonic black hole is actually a degenerate
thermodynamical system and there are no enough parameters to hold completely fixed when
performing the differential operations in the last subsection. Once taking into account
this chirality condition physically, the differential first law (\ref{FL}) and the integral
Bekenstein-Smarr formula (\ref{Smarr}) should be constrained by Eq. (\ref{const}), and
actually depict a degenerate thermodynamical system.

Choosing $J$ as a redundant variable (although it is a real observable quantity)\footnote{
Alternately, one can try to eliminate $\mathcal{P}$ rather than $J$ via Eq. (\ref{const})
also.} and eliminating it from the differential and integral mass formulae with the help
of $l^2 = 3/(8\pi\mathcal{P})$, the first law (\ref{FL}) and the Bekenstein-Smarr relation
(\ref{Smarr}) now boil down to the following nonstandard forms
\be
\begin{aligned}
(1 -\Omega\, l)dM &= TdS +V^{\prime} d\mathcal{P} +\Phi\, dQ +\Psi\, dP +K d\mu \, , \\
(1 -\Omega\, l)M &= 2(TS -V^{\prime}\mathcal{P}) +\Phi\, Q  +\Psi\, P \, ,
\end{aligned}
\ee
where
\bea
V^{\prime} = V -\frac{J\Omega}{2\mathcal{P}} = V -\frac{4\pi}{3}\Omega\, Ml^3 \, . \nn
\eea
It is clear that the thermodynamic quantities in the above formulae cannot comprise the
ordinary canonical conjugate pairs due to the existence of a factor $(1 -\Omega\, l)$ in
front of $dM$ and $M$.

Meanwhile, the squared-mass formula (\ref{sqm}) reduces to
\be\label{rMs}
M^2\Big(1 -\frac{\mu}{16\pi\mathcal{P}S}\Big) = \frac{8\mathcal{P}S}{3\mu}
 \bigg[\frac{4\mathcal{P}}{3}S^2 +\pi\big(P^2 +Q^2\big)\bigg] \, .
\ee

In this way, one actually regards the enthalpy $M$ as the fundamental functional relation
$M = M(S, Q, P, \mathcal{P}, \mu)$. Resembling the strategy employed in the last subsection,
one can deduce the above nonstandard differential and integral mass formulas from Eq.
(\ref{rMs}) by using the standard Maxwell rule.

\subsection{Horizon geometry and conformal boundary}

From now on, we shall concentrate on other basic properties, such as the horizon geometry
and conformal boundary, and RII of the ultraspinning dyonic Kerr-Sen-AdS$_4$ black hole,
as well as bounds on the mass and horizon radius in the extremal case. It is suggestive
to recast the metric (\ref{SEKSAdSdy}) into another helpful form
\bea
ds^2 &=& -\, \frac{\Delta(r)\Sigma\, dt^2}{\big[2m(r-d)-p^2-q^2\big]l^2}
 +\frac{\Sigma\, dr^2}{\Delta(r)} +\frac{\Sigma\, d\theta^2}{\sin^2\theta} \nn \\
&& +\frac{2m(r-d)-p^2-q^2}{\Sigma}\bigg\{l\sin^2\theta\, d\varphi -dt \nn \\
&& +\frac{\big(r^2-2dr-k^2+l^2\big)\Sigma}{\big[2m(r-d)-p^2-q^2\big]l^2}dt\bigg\}^2 \, .
\eea
To ensure that the spacetime outside the event horizon has the correct Lorentzian signature,
the following inequalities must be simultaneously satisfied:
\be
\begin{aligned}
&\Sigma \geq 0 \, , \qquad \Delta(r) \geq 0 \, , \\
& 2m(r-d) -p^2-q^2 = 2m\big(r -p^2/m\big) \geq 0 \, .
\end{aligned}
\ee
Given that the mass parameter of the ultraspinning dyonic black hole is absolutely positive
($m > 0$), it is immediately required that
\be\label{con1}
r \geq p^2/m \, .
\ee
And this also meets the requirement: $\Sigma = \big(r+q^2/m\big)\big(r-p^2/m\big) +l^2
\cos^2\theta \geq 0$. Then it can be checked that $g_{\varphi\varphi} = 2ml^2\big(r
-p^2/m\big)\sin^4\theta/\Sigma \geq 0$ is strictly guaranteed outside the event horizon,
and thus the geometry is free of any closed timelike curve (CTC). Finally, the condition
$\Delta(r) \geq 0$ results in the following inequalities:
\be\label{con2}
\big[\big(r+q^2/m\big)\big(r-p^2/m\big) +l^2\big]^2 \geq 2m\big(r-p^2/m\big)l^2 \geq 0 \, .
\ee
Only when the above two requirements (\ref{con1}) and (\ref{con2}) are meet, the spacetime
outside the event horizon is Lorentzian and free of CTC.

To explore the geometry of the event horizon, let us study the constant ($t, r$) surface on
which the induced metric is
\be\label{hm}
ds^2_h = \frac{\Sigma_+}{\sin^2\theta}d\theta^2 +\frac{\big(r_+^2
 -2dr_+ -k^2 +l^2\big)^2}{\Sigma_+}\sin^4\theta\, d\varphi^2 \, ,
\ee
where $\Sigma_+ = r_+^2 -2dr_+ -k^2 +l^2\cos^2\theta$. It is clear that this metric is singular
at $\theta = 0$ and $\theta = \pi$. Let us first examine whether the metric is ill-defined at
$\theta = 0$, and analyze it in the limit: $\theta\rightarrow 0$. In the small angle case
($\theta \sim 0$), we can introduce a new variable: $\kappa = l(1 -\cos\theta)$. Using
$\sin^2\theta \simeq 2\kappa/l$, the two-dimensional cross section (\ref{hm}) for small
$\kappa$ can be written as
\be\label{hm1}
ds^2_h = \big(r_+^2 -2dr_+ -k^2 +l^2\big)\bigg(\frac{d\kappa^2}{4\kappa^2}
 +\frac{4\kappa^2}{l^2}d\varphi^2\bigg) \, .
\ee
The above metric (\ref{hm1}) naturally reduces to what was considered \cite{PRD102-044007}
in the ultraspinning Kerr-Sen-AdS$_4$ black hole case when the magnetic charge parameter
vanishes ($p = 0$), and is clearly a metric of constant, negative curvature on a quotient
of the hyperbolic space $\mathbb{H}^2$. Due to the symmetry, one can perform a similar
analysis in the $\theta \sim \pi$ case and get the same result in the $\theta\rightarrow
\pi$ limit. Apparently, the space is free from pathologies near the north and south poles.
Topologically, the event horizon is a sphere with two punctures, and sometimes is called
the black spindle \cite{1912.03974}. This indicates that the above ultraspinning dyonic
Kerr-Sen-AdS$_4$ black hole owns a finite area but noncompact horizon.

Next, we wish to study the conformal boundary of the ultraspinning dyonic Kerr-Sen-AdS$_4$
black hole. After multiplying the metric (\ref{SEKSAdSdy}) with the conformal factor $l^2/r^2$
and taking the $r\to\infty$ limit, the boundary metric reads
\be\label{bdry}
ds_b^2 = -dt^2 +2l\sin^2\theta\, dtd\varphi +l^2d\theta^2/{\sin^2\theta} \, ,
\ee
which is the same one as those of the superentropic Kerr-Newman-AdS$_4$ black hole
\cite{JHEP0615096} and the ultraspinning Kerr-Sen-AdS$_4$ black hole \cite{PRD102-044007}.
Obviously, the coordinate $\varphi$ is null on the conformal boundary. In the small
$\kappa = l(1 -\cos\theta)$ limit, the conformal boundary metric (\ref{bdry}) can be
reexpressed as
\be
ds_b^2 = -dt^2 +4\kappa\, dtd\varphi +l^2\, d\kappa^2/\big(4\kappa^2\big) \, ,
\ee
which is interpreted in \cite{PRD89-084007} as an AdS$_3$ written as a Hopf-like fibration
over $\mathbb{H}^2$. It follows that the metric has nothing pathological near two poles:
$\theta = 0$ and $\theta = \pi$.

\subsection{Bounds on the mass and horizon radii
of extremal ultraspinning dyonic black holes}

In Ref. \cite{PRD102-044007}, we have discussed the bounds on the mass and horizon radius of
the extremal ultraspinning Kerr-Sen-AdS$_4$ black hole. Here we would like to seek some similar
inequalities for its dyonic counterpart. In the extremal dyonic black hole case, two roots of
the horizon equation $\Delta(r) = 0$ coincide with each other, and its location is determined
by $\Delta(r_e) =\Delta^{\prime}(r_e) = 0$, whose explicit expressions are given by
\be
\begin{aligned}
& \bigg(r_e^2 +\frac{q^2-p^2}{m}r_e -\frac{q^2p^2}{m^2} +l^2\bigg)^2
 = 2ml^2\bigg(r_e -\frac{p^2}{m}\bigg) \, , \\
& \bigg(r_e^2 +\frac{q^2-p^2}{m}r_e -\frac{q^2p^2}{m^2} +l^2\bigg)
 \bigg(2r_e +\frac{q^2-p^2}{m}\bigg) = ml^2 \, , \quad
\end{aligned}
\ee
from which one can get a quadratic equation and a cubic equation about the radius $r_e$:
\bea
r_e^2 +\frac{q^2 -5p^2}{3m}r_e -\frac{m^2l^2
 +p^2\big(q^2 -2p^2\big)}{3m^2} &=& 0 \, , \label{r2} \\
\bigg(r_e -\frac{p^2}{m}\bigg)\bigg(r_e
 +\frac{q^2 -p^2}{2m}\bigg)^2 -\frac{ml^2}{8} &=& 0 \, .
\label{r3}
\eea
Using Eq. (\ref{r2}), one can eliminate the $r_e^3$ and $r_e^2$ terms from Eq. (\ref{r3})
to arrive at the expression for the extremal horizon radius:
\be
r_e = \frac{p^2}{m} -8ml^2\frac{p^2+q^2 -9m^2/16}{\big(p^2+q^2\big)^2 +12m^2l^2} \, ,
\ee
and resubmit it into Eqs. (\ref{r2}) and (\ref{r3}) to obtain an important equality
relating the solution parameters:
\bea
&&\bigg[l^2+\frac{27}{256}m^2 -\frac{1}{4m^2}\Big(p^2+q^2 -\frac{9}{8}m^2\Big)^2\bigg]^2 \nn \\
&&\qquad = -\frac{1}{4m^2}\Big(p^2+q^2-\frac{9}{16}m^2\Big)^3 \, , \label{pse}
\eea
which will give a stringent restriction on the parameters range allowed by the extremal
dyonic solution.

In order to analyze the parameters equation (\ref{pse}), it is convenient to include two new
variables: the rescaled mass $y = m/l$ and the rescaled scalar charge $x = (p^2+q^2)/(2ml) =
\sqrt{d^2+k^2}/l$ to rewrite Eq. (\ref{pse}) as a quadratic equation of $y$:
\be
\frac{27}{64}y^2 +\frac{1}{4}x\big(x^2-9\big)y -\big(x^2-1\big)^2 = 0 \, ,
\ee
which admits two real roots:
\be
y^{\pm} = \frac{8}{27}\Big[x\big(9-x^2\big) \pm \big(x^2+3\big)^{3/2}\Big] \, .
\ee
Meanwhile, we would like to introduce also two shifted radii\footnote{This suggests to
adopt the radial coordinate: $\rho = r -p^2/m$ or $R = r -d$.}
\be
\begin{aligned}
\rho_e &= r_e -\frac{p^2}{m} = \frac{9y-32x}{8\big(x^2+3\big)}l \, , \\
R_e &= r_e -d  = l\, x +\frac{9y-32x}{8\big(x^2+3\big)}l \, . \quad
\end{aligned}
\ee

\begin{figure}[!htbp]
\centering
\subfigure[\, Scaled mass vs rescaled scalar charge]{\label{sfa}
\includegraphics[width=0.45\textwidth,height=0.225\textheight]{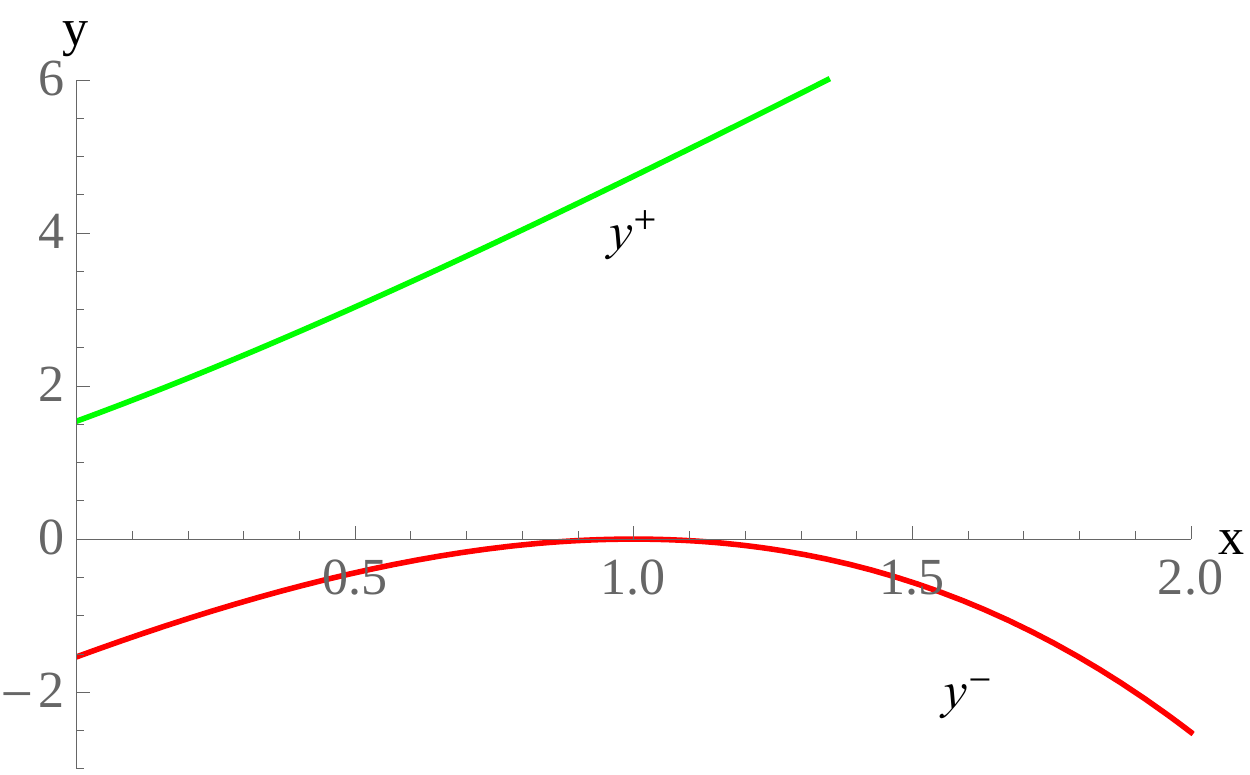}}
\hspace{1cm}
\subfigure[\, Shifted radii vs rescaled scalar charge with $l=1$]{\label{sfb}
\includegraphics[width=0.45\textwidth,height=0.225\textheight]{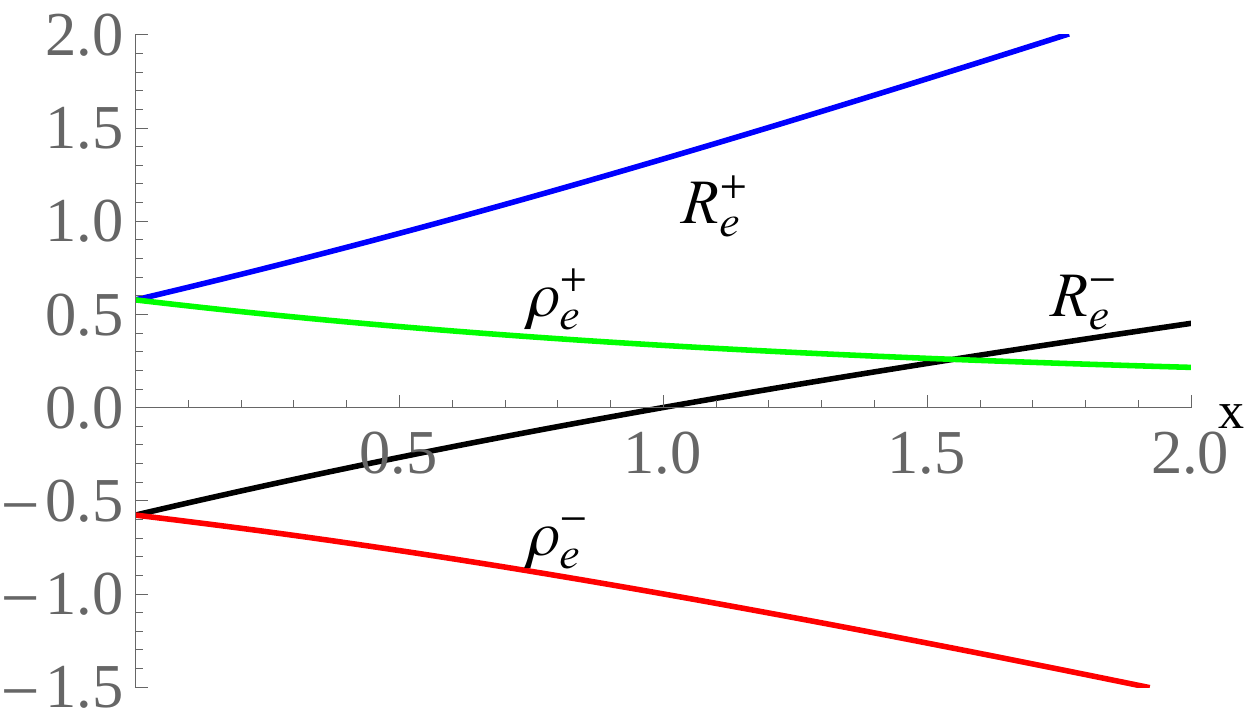}}
\caption{On the basis of the reasonable range of the positive mass and cosmological scale
($m>0$ and $l>0$), the negative root $y^-$, and accordingly $\rho_e^-$ and $R_e^-$ should
be excluded from our discussions on the ground of physical reason. In the interval $x\in
[0, \infty)$, the root $y^+$ is a monotonic increasing function of $x$, $\rho_e^+$ is
monotonic decreasing but $R_e^+$ is monotonic increasing with increasing $x$. The origin
$x = 0$ is equivalent to $p = q = 0$, which corresponds to the extremal ultraspinning
Kerr-AdS$_4$ black hole case.}
\label{sf}
\end{figure}

In Fig. \ref{sf}, we plot the rescaled mass and rescaled shifted radii as functions of the
rescaled scalar charge in the physical range $x\in [0, \infty)$. For the positive mass and
cosmological scale ($m>0$ and $l>0$), only the root $y^+$ is admissible, and acquires a
minimal value: $8\sqrt{3}/9$ at $x = 0$ ($p = q =0$). At the same time, $\rho_e^+$ and
$R_e^+$ intersect at the Hawking-Page phase transition scale: $r_{\rm HP} = l/\sqrt{3}$
when $x = 0$. This implies that the extremal mass has the lowest bound:
\be
m = m_e \geq \frac{8l}{3\sqrt{3}} \, ,
\ee
but $r_{\rm HP} = l/\sqrt{3}$ becomes, respectively, the upper and lower bounds of the
shifted horizon radii:
\be
\rho_e^+ \leq \frac{l}{\sqrt{3}} \leq  R_e^+ \, .
\ee
The above bounds on the extremal mass and horizon radii are in complete accordance with
the results previously obtained in the extremal ultraspinning Kerr-Sen-AdS$_4$ case
\cite{PRD102-044007}. In particular, the detailed discussion made in this subsection
further confirms the conclusion in the Note-added of our previous paper \cite{PRD102-044007}.

\subsection{Reverse isoperimetric inequality}

It has been conjectured \cite{PRD84-024037} that the AdS black holes fulfil the following RII:
\be
\mathcal{R} = \bigg[\frac{(D-1)V}{\cA_{D-2}}\bigg]^{1/(D-1)}
 \bigg(\frac{\cA_{D-2}}{A}\bigg)^{1/(D-2)} \geq 1 \, ,
\ee
with $\cA_{D-2} = 2\pi^{[(D-1)/2]}/\Gamma[(D-1)/2]$ being the area of the unit $(D-2)$-sphere
and $A = 4S$ the horizon area. Equality is attained for the Schwarzschild-AdS black hole, which
implies that the Schwarzschild-AdS black hole has the maximum entropy. In other words, for a
given entropy, the Schwarzschild-AdS black hole owns the least volume.

Now, we would like to directly check whether or not the ultraspinning dyonic Kerr-Sen-AdS$_4$
black hole obeys this RII. We have already known that the area of the unit two-dimensional
sphere, the thermodynamic volume, and the horizon area are: $\mathcal{A}_2 = 2\mu$,  $V =
4(r_+ -d)S/3$, and $A = 4S = 2\mu\big(r_+^2 -2dr_+ -k^2 +l^2\big)$, respectively. Therefore,
the isoperimetric ratio is
\bea
\mathcal{R} &=& \bigg(\frac{r_+ -d}{2\mu}A\bigg)^{1/3}\bigg(\frac{2\mu}{A}\bigg)^{1/2} \nn \\
&=& \bigg[\frac{(r_+ -d)^2}{(r_+ -d)^2 -d^2 -k^2 +l^2}\bigg]^{1/6} \, .
\eea
Clearly, the ratio of $\mathcal{R}$ is uncertain. If $0\leq d^2+k^2 < l^2$ (namely, $0\leq
p^2 +q^2 < 2ml$), then $\mathcal{R} < 1$, which implies that the ultraspinning dyonic
Kerr-Sen-AdS$_4$ black hole violates the RII, and is superentropic. Otherwise if $d^2 +k^2
\geq l^2$ (or $p^2 +q^2 \geq 2ml$), one then obtains $\mathcal{R} \geq 1$. In this case, the
ultraspinning dyonic Kerr-Sen-AdS$_4$ black hole obeys the RII, and is subentropic. Because
the value range of $\mathcal{R}$ crucially depends upon the values of the solution parameters
($p$, $q$, $m$ and $l$), one can find that the ultraspinning dyonic Kerr-Sen-AdS$_4$ black hole
is not always superentropic, similar to the purely electric-charged case that describes the
ultraspinning Kerr-Sen-AdS$_4$ black hole \cite{PRD102-044007}. Only when the parameters obey
the inequality $p^2 +q^2 < 2ml$ does it violate the RII, whilst the superentropic dyonic
Kerr-Newman-AdS$_4$ black hole always violates the RII \cite{PRL115-031101}. As far as this
point is concerned, these dyonic AdS$_4$ black holes exhibit one remarkable different property.

\section{Conclusions}\label{V}

In this paper, we have extended our previous work \cite{PRD102-044007} to a more general dyonic
case and investigated some interesting properties of the dyonic Kerr-Sen-AdS$_4$ black hole and
its ultraspinning counterpart in the four-dimensional gauged EMDA theory. To this end, we first
presented an exquisite form for the dyonic Kerr-Sen black hole solution and found its generalization
by including a nonzero negative cosmological constant, namely, the dyonic Kerr-Sen-AdS$_4$ black
hole. Then by applying a simple $a\to l$ limit procedure, we obtained its ultraspinning cousin.
All the expressions of these solutions, namely, their metric, the Abelian gauge potential and its
dual potential, as well as the dilaton scalar and axion pseudoscalar fields are very convenient
for exploring their thermodynamical properties. We presented all necessary thermodynamic quantities
and demonstrated that they obey both the differential and integral mass formulae. Furthermore,
we displayed new Christodoulou-Ruffini-like squared-mass formulae for these four-dimensional
AdS$_4$ black holes, from which all expected thermodynamic conjugate partners can be computed
by differentiating these squared-mass formulae with respect to their corresponding thermodynamic
variables and are demonstrated to constitute the ordinary canonical conjugate pairs in the
standard forms of black hole thermodynamics.

In particular, we have utilized the method advocated in Refs. \cite{PRD101-024057,PRD102-044007}
to show that all thermodynamical quantities of the ultraspinning dyonic Kerr-Sen-AdS$_4$ black
hole can be obtained via taking the same ultraspinning $a\to l$ limit to those of their
corresponding predecessor in the rotating frame at infinity. After that, we have discussed
in detail about the impact of the chirality condition on the actual thermodynamics of this
ultraspinning dyonic black hole. To a certain extent, these aspects resemble those of the
superentropic dyonic Kerr-Newman-AdS$_4$ and the ultraspinning Kerr-Sen-AdS$_4$ black hole.

Paralleling to the work done in Ref. \cite{PRD102-044007}, we have discussed some bounds on
the mass and horizon radius of the extremal ultraspinning dyonic Kerr-Sen-AdS$_4$ black hole.
Our results further confirm those established for the ultraspinning Kerr-Sen-AdS$_4$ black
hole \cite{PRD102-044007}, and reproduce its conclusion when the magnetic charge parameter
vanishes.

Like the purely electric-charged case of the ultraspinning Kerr-Sen-AdS$_4$ black hole
\cite{PRD102-044007}, we have also found that the ultraspinning dyonic Kerr-Sen-AdS$_4$ black
hole is not always superentropic, since the RII is violated only when $p^2 +q^2 < 2ml$. Once
$p^2 +q^2 \geq 2ml$, the ultraspinning dyonic Kerr-Sen-AdS$_4$ black hole will be subentropic.
This black hole  resembles  the ultraspinning Kerr-Sen-AdS$_4$ black hole which does not always
violate the RII \cite{PRD102-044007}, but is in sharp contrast with the superentropic dyonic
Kerr-Newman-AdS$_4$ black hole that always violates the RII \cite{PRL115-031101}. A most related
issue is to investigate whether or not the RII is violated in the reduced form of extended
thermodynamic phase space, as did in Ref. \cite{PLB807-135529}.

\acknowledgments

This work is supported by the National Natural Science Foundation of China under Grants No.
11675130, No. 11275157, No. 11775077, No. 12075084, No. 11690034, and No. 11435006, and by
the Science and Technology Innovation Plan of Hunan province under Grant No. 2017XK2019.
We are greatly indebted to the anonymous referee for his/her invaluable comments and good
suggestions to improve the presentations of this work.

\end{document}